\documentclass[%
 reprint,
 superscriptaddress,
 amsmath,amssymb,
 aps,
 prl,
]{revtex4-2}

\usepackage{physics}
\usepackage{graphicx}
\usepackage{dcolumn}
\usepackage{bm}
\usepackage{booktabs} 
\usepackage{xcolor}
\definecolor{deeppurple}{HTML}{3F007D} 
\usepackage{hyperref}
\hypersetup{colorlinks=true,allcolors=deeppurple}

\begin{document}

\author{Jonas H{\"a}nseroth}
\email{jonas.haenseroth@tu-ilmenau.de}
\affiliation{Theoretical Solid State Physics, Institute of Physics, Technische Universit{\"a}t Ilmenau, 98693 Ilmenau, Germany}

\author{Christian Dre{\ss}ler}
\affiliation{Theoretical Solid State Physics, Institute of Physics, Technische Universit{\"a}t Ilmenau, 98693 Ilmenau, Germany}

\title{Data-Efficient Construction of Material-Specific Machine-Learning Interatomic Potentials from Ab Initio Molecular Dynamics Trajectories}

\date{\today}

\begin{abstract}
Large-scale pretrained machine-learning interatomic potentials, so-called universal or foundation models provide an appealing starting point for atomistic simulations, but their accuracy for material-specific observables often remains limited without additional reference data (fine-tuning). 
Here, we systematically quantify how much first-principles data are required to convert universal models into ab initio-accurate material-specific potentials and we ask whether fine-tuning is necessarily preferable to training models from scratch.
We compare five universal MLIP frameworks, \textsc{MACE-MP-0} (small), \textsc{SevenNet-0}, \textsc{GRACE-1L-OAM}, \textsc{MatterSim-v1.0.0-5M} and \textsc{ORB-v2}, across seven chemically diverse systems incorporating rare and reactive events. 
Fine-tuning on only $10$~AIMD-derived configurations is for the investigated systems insufficient; $200$~configurations succeed in favorable cases, but the outcome remains strongly system-dependent. 
By contrast, $2{,}000$~AIMD configurations constitute a robust default, yielding low force and energy errors and reproducing the investigated material-specific observables.
Moderately dense sub-sampling of the AIMD trajectory reduces the required trajectory length tenfold without substantially degrading model quality.
Training-from-scratch on the same datasets is competitive with and often slightly more accurate than, naive fine-tuning for \textsc{MACE} and \textsc{SevenNet}, whereas \textsc{GRACE} requires more data in this setting. 
The energy profile for a sulfur-vacancy jump in MoS$_2$ demonstrates that low trajectory-level errors do not necessarily guarantee a correct reaction energy profile, highlighting the need for observable-level validation. 
Finally, we show that averaging independently trained models improves predictions in scarce-data regimes at no additional first-principles cost.
Together, these results provide practical guidelines for converting limited AIMD reference data into reliable material-specific MLIPs for nanosecond-timescale simulations at near-DFT accuracy.
\end{abstract}

\maketitle

\section*{Introduction}

The usefulness of an atomistic simulation is ultimately determined by whether it reproduces the physical observable of interest.
For many materials problems, this observable is not controlled by the most frequently sampled equilibrium configurations, but by rare or collective regions of configuration space.
Diffusion barriers, proton-transfer pathways, defect migration, hydrogen-bond rearrangements, solvation structures and bond-breaking events all depend on subtle features of the potential-energy surface that are difficult to sample and even more difficult to model reliably \cite{grunert2025,dressler2020effect,qaisrani2025bridging,kirsch2022atomistic}.
This poses a central challenge for computational materials science: simulations must be long enough to observe the relevant process a statistically meaningful number of times, yet accurate enough to describe the local chemistry that governs it.

Density-functional-theory-based (DFT) ab initio molecular dynamics (AIMD) provides the required level of chemical fidelity because the electronic structure is evaluated explicitly along the trajectory \cite{marx2000ab,tuckerman2002ab,iftimie2005ab}.
However, the cost of repeated first-principles calculations restricts AIMD to limited system sizes and simulation times.
Classical force fields overcome this limitation and enable much longer trajectories, but their fixed functional forms and parameterization domains often preclude reliable extrapolation to reactive, defective, or chemically complex environments \cite{plimpton1995computational,sutmann2002classical,brooks2021classical}.
Consequently, many problems that demand both long-time sampling and near-quantum accuracy remain inaccessible to either approach alone.

Machine-learning interatomic potentials (MLIPs) have transformed this landscape by shifting much of the computational effort from the simulation itself to the generation of reference data.
Once trained on quantum-mechanical energies and forces, a MLIP evaluates the potential-energy surface at a cost much closer to that of classical molecular dynamics while retaining near ab initio accuracy within the domain covered by the training data \cite{wang2020machine,pravsnikar2024machine,kabylda2025molecular,poltavsky2025crash}.
Early neural-network potentials and Gaussian approximation potentials established this principle for carefully constructed datasets \cite{behler2007,bartok2010,friederich2021}.
Subsequent developments in graph neural networks, equivariant message passing and symmetry-preserving representations have further improved data efficiency, accuracy and stability across molecular and materials systems \cite{thomas2018tensor,batzner20223,mace_1,mace_2,unke2021,reiser2022,grace_1,drautz2019}.

A recent step has been the construction of so-called universal or foundation MLIPs.
These models are not optimized for a single target material but are pretrained on large and chemically diverse collections of DFT-labeled structures \cite{jacobs2025practical,mace_mp,mattersim,orbv3,grace_2}, drawn from large-scale repositories and atomistic databases such as the \textsc{Materials Project}, \textsc{Alexandria}, \textsc{Open Materials 2024} and \textsc{Open Molecules 2025} \cite{alexandria,mp_1,mp_2,omat24,omol25}.
Frameworks such as \textsc{MACE}, \textsc{GRACE}, \textsc{SevenNet}, \textsc{MatterSim} and \textsc{ORB} have thereby made it possible to apply a single pretrained potential across a broad chemical space \cite{mace_mp,grace_2,mattersim,sevennet_2,orbv3}.
Their strong performance in broad benchmarks, including \textsc{Matbench Discovery} and \textsc{MLIP Arena}, has made them attractive both as general-purpose models and as starting points for material-specific simulations \cite{matbench,chiang2025mlip}.

Broad chemical coverage, however, does not automatically translate into quantitative reliability for a particular material property.
A universal model may achieve favorable average force and energy errors across many structures while still failing in the local region of configuration space that controls a specific reaction coordinate, diffusion event, or phase-dependent mechanism.
This distinction is especially important for rare events and high-barrier processes, whose relevant configurations are sampled infrequently and may be underrepresented in the pretraining data \cite{grunert2025,haenseroth2026htscreening,flototto2026large,haenseroth2026atk,haenseroth2026umlip_config_space}.
Recent studies have therefore emphasized that universal MLIPs often require system-specific adaptation before they can be used for quantitative simulations of specialized materials properties \cite{grunert2025,flototto2026large,hanseroth2025optimizing,weiske2025statistics,chen2025high,liu2025fine,kaur2025data,haenseroth2026revealing}.

The most common adaptation strategy is fine-tuning, in which the parameters of a pretrained universal model are updated using a smaller DFT-labeled dataset of the target system, allowing the model to correct its potential-energy surface in the region most relevant to the intended application \cite{grunert2025,flototto2026large,hanseroth2025optimizing,weiske2025statistics,chen2025high,liu2025fine,kaur2025data,haenseroth2026revealing,radova2025fine, tompa2026fine}.
Several variants have been proposed to balance the retention of general chemical knowledge against adaptation to the new system.
Frozen-layer fine-tuning updates only selected network components while keeping the remaining representation fixed \cite{radova2025fine}.
Low-rank adaptation methods, such as LoRA, introduce a small number of trainable parameters to reduce memory requirements and improve parameter efficiency \cite{hu2022lora,grandel2026parameter}.
Multi-head replay strategies retain a shared representation while using separate output heads or replayed datasets to mitigate forgetting during adaptation \cite{mace_1,sevennet_2,beck2025multihead}.

Despite the growing use of fine-tuned universal MLIPs, practical guidance remains incomplete.
In particular, the amount of DFT-labeled data needed for reliable material-specific accuracy is still unclear.
Very small datasets are attractive because they minimize the cost of reference calculations, but they may not contain enough information to correct the potential-energy surface in physically important regions.
Larger datasets are more robust, but their generation can become the dominant cost of the workflow.
The optimal compromise depends not only on the number of configurations, but also on how those configurations are sampled from the underlying dynamics.

A second unresolved question is whether fine-tuning is always preferable to training models from scratch.
universal models provide chemically informed initialization, but a smaller material-specific model trained directly on the same DFT-labeled configurations may be competitive once enough target-system data are available.

In this work, we systematically address both questions for material-specific MLIPs constructed from AIMD-derived reference datasets.
Rather than focusing on advanced adaptation strategies, we deliberately compare two simple and widely accessible approaches: naive fine-tuning of universal models and training models from scratch on the same DFT-labeled configurations.
This design isolates the roles of dataset size, trajectory sub-sampling and model initialization.
For fine-tuning, we consider five prominent universal MLIP frameworks: \textsc{MACE}, \textsc{GRACE}, \textsc{SevenNet}, \textsc{MatterSim} and \textsc{ORB} \cite{mace_1,grace_1,sevennet_1,mattersim,orbv2}.
For training-from-scratch, we use model sizes matched to the corresponding universal models for \textsc{MACE}, \textsc{SevenNet} and \textsc{GRACE}, enabling a direct comparison between both routes to material-specific potentials.

The benchmark covers seven chemically diverse systems chosen to probe different bonding motifs, phases and dynamical processes:
cesium dihydrogen phosphate (CDP) and Cs$_7$(H$_4$PO$_4$)(H$_2$PO$_4$)$_8$ (CPP) as proton-conducting solid acid materials \cite{struct_cdp,struct_cpp,dressler2023coexistence}, L-pyroglutamate-ammonium (L-Pyro) as an organic crystal with low-barrier hydrogen bonds \cite{stephens2021short,miron2023carbonyl,qaisrani2025acid,qaisrani2025bridging,haenseroth2026nqedisp}, solvated phenol (PhOH) as a liquid-phase hydrogen-bonding system, aqueous potassium hydroxide (KOH) as a concentrated electrolyte \cite{hanseroth2025optimizing,haenseroth2025ohlmc}, crystalline Li$_{13}$Si$_4$ as an intermetallic lithium silicide \cite{zeilinger2013revision,kirsch2025li+,kirsch2022atomistic} and MoS$_2$ with sulfur vacancies as a defective two-dimensional material \cite{Li2018,Spetzler2024,flototto2026large}.
Together, these systems constitute a demanding test set for assessing whether MLIPs reproduce not only trajectory-level energy and force errors but also the target observables relevant to each material.

We first determine how the performance of fine-tuned universal MLIPs depends on the number of DFT-labeled AIMD configurations.
Datasets containing $10$, $200$ and $2{,}000$~structures allow us to distinguish between extremely data-scarce, low-data and more robust intermediate-data regimes.
We then investigate whether the training structures must originate from a long, sparsely sampled AIMD trajectory, or whether shorter trajectories sampled more densely provide comparable model quality.
This comparison directly addresses the practical cost of generating reference data, because denser sub-sampling shortens the required AIMD trajectory at the expense of increased correlations between consecutive trajectory configurations.

\begin{figure*}
    \centering
    \includegraphics{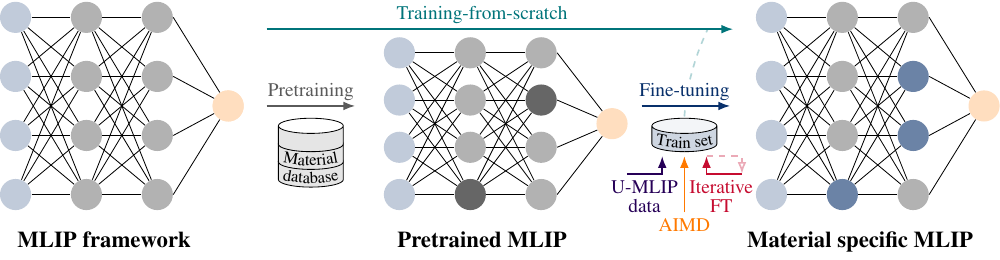}
    \caption{
    \textbf{MLIP training strategies.}
    Overview of routes from a general MLIP framework (e.g., \textsc{MACE}) to a material-specific MLIP. Strategies include training-from-scratch, fine-tuning pretrained universal models with AIMD-derived datasets, using structures sampled by universal models and recalculated with DFT and iterative fine-tuning workflows.
    }
    \label{fig:strategies}
\end{figure*}

We next compare these fine-tuned models with models trained-from-scratch on identical datasets (Fig.~\ref{fig:strategies}).
This comparison separates the value of the pretrained initialization from the value of the DFT-labeled target-system data itself.
In addition, we explore whether averaging several independently trained models improves predictions when the training dataset is limited.
Such model averaging requires no additional DFT calculations and therefore offers a simple means of increasing robustness in scarce-data regimes.

A key aspect of our analysis is that we do not rely exclusively on force and energy errors.
As an observable-level test, we evaluate the sulfur-vacancy jump in MoS$_2$ using nudged elastic band (NEB) calculations.
This example reveals whether a model recovers the correct potential-energy profile for a high-barrier defect-migration process and it exposes cases in which seemingly acceptable force and energy errors are insufficient to reproduce the relevant physical pathway \cite{fu2022forces}.
This distinction between numerical accuracy on trajectory data and physical accuracy for the target observable is central to the construction of reliable material-specific MLIPs.

The results lead to practical guidelines for building ab initio-accurate material-specific MLIPs from limited first-principles data and define a workflow in which limited AIMD reference data are converted into reliable material-specific MLIPs for nanosecond-scale simulations with near ab initio accuracy.

\section*{Results}

\subsection*{Dataset size}

\begin{figure*}
    \centering
    \includegraphics{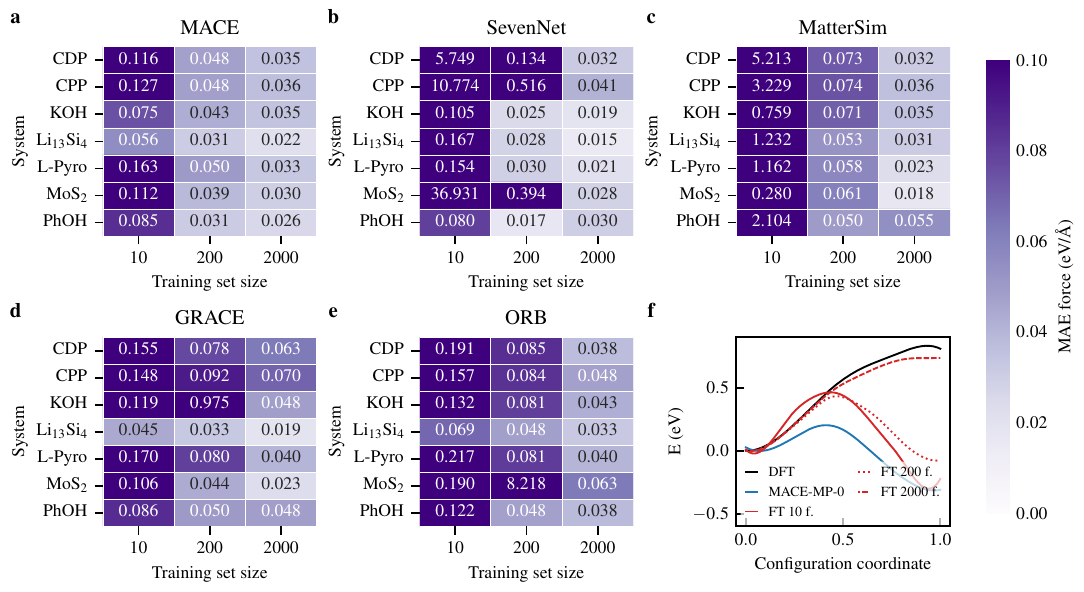}
    \caption{
    \textbf{Dataset-size influence for fine-tuned universal MLIPs.}
    Force mean absolute error in eV/{\AA} for different fine-tuned universal models (\textbf{a}~\textsc{MACE-MP-0} (small), \textbf{b}~\textsc{SevenNet-0}, \textbf{c}~\textsc{MatterSim-v1.0.0-5M}, \textbf{d}~\textsc{GRACE-1L-OAM} and \textbf{e}~\textsc{ORB-v2}) on the seven benchmark systems (CDP, CPP, KOH, Li$_{13}$Si$_4$, L-Pyro, MoS$_2$ and PhOH) for training sets containing $10$, $200$, or $2{,}000$~AIMD configurations sampled equidistantly using stride $100$.
    A corresponding comparison of energy errors is provided in the Supplementary Note~3.
    \textbf{f}~Potential-energy curves for a sulfur jump into a neighboring line of sulfur vacancies, obtained from nudged elastic band calculations using DFT (solid black line), \textsc{MACE-MP-0} universal model (solid blue line), \textsc{MACE-MP-0} models fine-tuned on $10$ configurations (solid red line), $200$~configurations (dotted red line) and $2{,}000$~configurations (dashed red line).
    }
    \label{fig:dataset_size_ft}
\end{figure*}

Reactive events and high-barrier processes place particularly stringent demands on machine-learned interatomic potentials.
In our previous analysis, universal MLIPs often failed to predict the relevant physical observables with the accuracy required for such processes \cite{haenseroth2026atk, haenseroth2026umlip_config_space}.
A natural route to overcome this limitation is to supplement a universal model with a small system-specific reference dataset ($\leq0.1$~\% of the size of the pretraining set).
Here, we therefore ask how many first-principles data points are needed to turn a general-purpose MLIP into a reliable material-specific model.

We evaluated this question for five universal MLIP frameworks across seven chemically diverse materials.
Each model was fine-tuned on datasets containing $10$, $200$, or $2{,}000$~configurations sampled from AIMD trajectories.
The $10$-configuration datasets serve as an intentionally data-scarce lower bound rather than a realistic production strategy, whereas the $200$- and $2{,}000$-configuration datasets represent low- and intermediate-data regimes, respectively.
Force errors of the fine-tuned models are shown in Figure~\ref{fig:dataset_size_ft}a-e while the corresponding performance of the universal models is presented in Supplementary Notes~1 and 2. 
The corresponding energy errors of the fine-tuned models are reported in Supplementary Note~3.

Fine-tuning on only $10$~configurations is generally insufficient for production-quality models.
Although most models match or slightly improve upon their universal counterparts, the resulting force and energy errors remain too large in many cases and the target physical observables are not reproduced consistently.
This behavior indicates that a very small number of reference structures can locally correct a universal model but does not provide enough information to define a reliable material-specific potential-energy surface.

The $200$-configuration regime is more useful, but its success remains strongly system-dependent.
For some combinations of MLIP framework and material, in particular Li$_{13}$Si$_4$, L-Pyro and PhOH, the resulting force errors already fall within a practically useful range.
In contrast, CDP, CPP, KOH and MoS$_2$ remain challenging, with force errors frequently exceeding $0.1$~eV/{\AA}.
Energy errors follow the same qualitative trend and often remain comparatively large.
Thus, $200$~configurations can suffice for favorable systems, but this dataset size does not provide a generally reliable route to ab initio-quality material-specific MLIPs.

Fine-tuning on $2{,}000$~configurations is substantially more robust.
Across the investigated systems and MLIP frameworks, force errors are typically below $0.05$~eV/{\AA} and frequently below $0.03$~eV/{\AA}, while energy errors are mostly below $0.003$~eV/atom and often below $0.001$~eV/atom.
Within the scope of the present benchmark, $2{,}000$~DFT-labeled configurations therefore constitute a practical default dataset size when the goal is a material-specific MLIP with ab initio accuracy.

The sulfur-vacancy jump in MoS$_2$ provides a more stringent test than trajectory-level errors alone.
Figure~\ref{fig:dataset_size_ft}f shows the energy profile predicted by \textsc{MACE-MP-0} models fine-tuned on the different dataset sizes.
The universal \textsc{MACE-MP-0} model shows a high force error of $0.510$~eV/{\AA}.
The model fine-tuned on $200$~configurations reaches force and energy errors of $0.039$~eV/{\AA} and $0.0026$~eV/atom, respectively, yet still fails to reproduce the DFT energy profile.
Its energy curve resembles that of the model fine-tuned on only $10$~configurations, although both lie closer to DFT than the universal model in the first half of the reaction coordinate.
Only the \textsc{MACE-MP-0} model fine-tuned on $2{,}000$~configurations recovers the DFT energy profile.
The same qualitative behavior is observed for the other MLIP frameworks (Supplementary Note~4): models fine-tuned on $10$ or $200$~configurations do not reproduce the sulfur-vacancy jump, whereas models trained on $2{,}000$~configurations are substantially more reliable.
The material-specific properties predicted for the other chemical systems are shown in Supplementary Note~5.

\subsection*{Trajectory sub-sampling}

So far, all fine-tuning datasets were generated by equidistant sub-sampling of AIMD trajectories using stride $100$.
Because one of our aims is to reduce the number of first-principles calculations, we next asked whether a shorter AIMD trajectory sampled more densely provides comparable performance.
For a fixed dataset size, reducing the stride from $100$ to $10$ shortens the required AIMD trajectory by a factor of ten, but it also increases the similarity between neighboring training structures and narrows the sampled configurational space.
The force and energy errors of the five MLIP frameworks fine-tuned on stride-$10$ datasets for all seven materials are shown in Supplementary Notes~6 and~7, respectively. 

The stride-$10$ and stride-$100$ datasets exhibit the same overall trends.
Errors decrease with increasing dataset size and the relative behavior of the different MLIP frameworks remains similar, as do the absolute errors.
This indicates that, for the systems studied here, the number of training configurations is often more important than the total length of the AIMD trajectory from which they were sampled.
In particular, a dataset of $2{,}000$~configurations generated using stride $10$ can outperform a dataset of $200$~configurations generated from a trajectory of the same length using stride $100$.
Moderately dense sub-sampling is therefore an efficient strategy for reducing the cost of reference-data generation, provided that the resulting configurations still cover the relevant local environments.

\begin{figure*}
    \centering
    \includegraphics{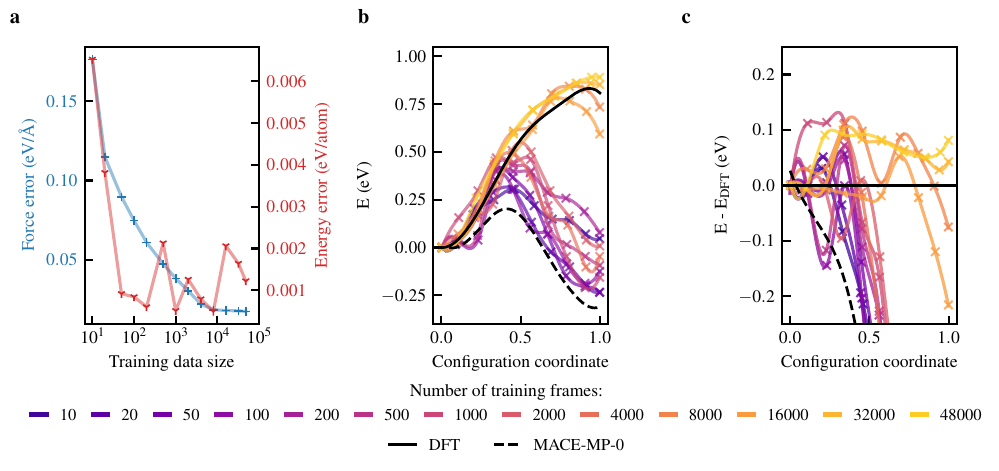}
    \caption{
    \textbf{Scaling behavior and error evaluation.}
    Scaling of fine-tuned MACE-MP-0 model performance with training-set size, from $10$ to $48{,}000$~AIMD-derived configurations of MoS$_2$ (stride~$1$), measured by force and energy errors (\textbf{a}), the predicted energy curve (\textbf{b}) and the deviation from the DFT energy curve (\textbf{c}).
    }
    \label{fig:scaling_law}
\end{figure*}

As an extreme limit, we also considered training sets constructed from every frame of an AIMD trajectory.
The resulting scaling behavior is shown in Figure~\ref{fig:scaling_law}a-c.
The force error decreases rapidly with increasing dataset size, starting above $0.15$~eV/{\AA} and approaching a plateau for datasets larger than approximately $8{,}000$~configurations.
The energy error is already small for the small datasets, typically in the range of $0.001$-$0.002$~eV/atom and changes less systematically with dataset size.
This weaker dependence of the energy error is expected because, for MoS$_2$, the energy loss was weighted $100$ times less than the force loss during fine-tuning.
The DFT energy profile is reproduced correctly by all investigated models trained on more than $4{,}000$~configurations, while the force error continues to decrease smoothly with dataset size (Figure~\ref{fig:scaling_law}).
These results show that dense sampling can eventually recover the correct barrier profile, but also that highly correlated data become increasingly redundant beyond the first few thousand configurations.

\subsection*{Training strategies}

\begin{figure}
    \centering
    \includegraphics[width=0.4\textwidth]{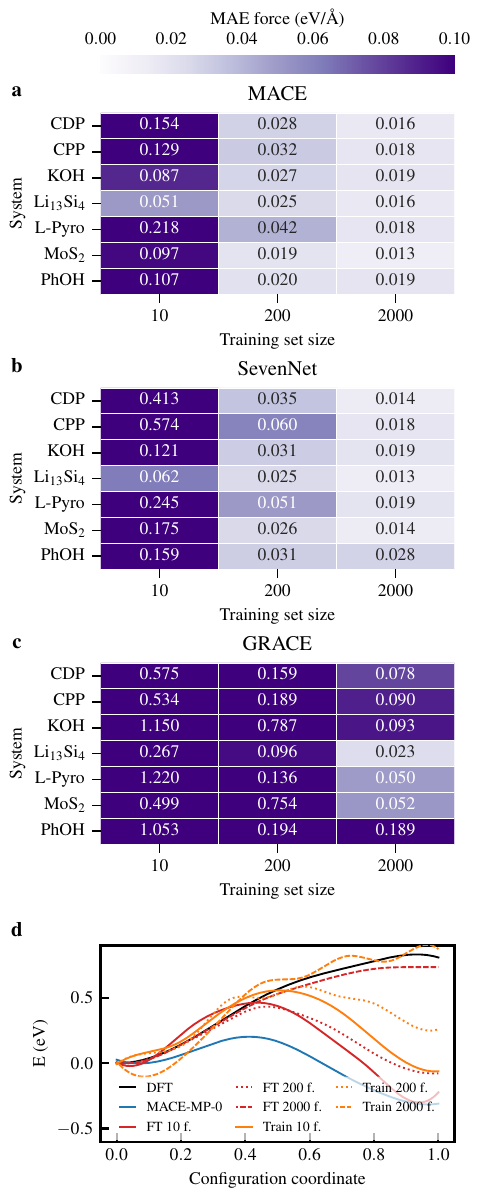}
    \caption{
    \textbf{Performance of MLIPs trained-from-scratch.}
    \textbf{a}-\textbf{c}~Force mean absolute error in eV/{\AA} for models trained-from-scratch with \textsc{MACE}, \textsc{SevenNet} and \textsc{GRACE} on the seven benchmark systems for training sets containing $10$, $200$, or $2{,}000$~AIMD configurations sampled equidistantly using stride $100$.
    \textbf{d}~Potential-energy curves for a sulfur jump into a neighboring line of sulfur vacancies, obtained from nudged elastic band calculations using fine-tuned (red lines) or trained-from-scratch (orange lines) \textsc{MACE-MP-0}-size models trained on $10$~configurations (solid lines), $200$~configurations (dotted lines), or $2{,}000$~configurations (dashed lines); the \textsc{MACE-MP-0} universal model (solid blue line); and DFT (solid black line).
    The corresponding energy errors are shown in Supplementary Note~8.
    } 
    \label{fig:train}
\end{figure}

Fine-tuning is not the only route to material-specific MLIPs.
An alternative is to train a model from scratch on the same system-specific reference data, without initializing from the weights of a pretrained universal model.
To compare the two strategies directly, we trained \textsc{MACE}, \textsc{SevenNet} and \textsc{GRACE} models from scratch using the same datasets as in the fine-tuning experiments.
For each framework, the model size was chosen to match the corresponding universal model used for fine-tuning: \textsc{MACE-MP-0} (small), \textsc{SevenNet-0} and \textsc{GRACE-1L-OAM}.

The models trained-from-scratch reproduce the same basic data-size trend observed for fine-tuning.
Increasing the dataset from $10$ to $200$~configurations substantially reduces the error and a further increase to $2{,}000$~configurations often yields an additional improvement.
However, the framework dependence is more pronounced than for fine-tuning.
Models trained-from-scratch with \textsc{GRACE} perform considerably worse than those trained with \textsc{MACE} or \textsc{SevenNet}, a trend that is also reflected in the corresponding energy errors reported in Supplementary Note~8.

For \textsc{MACE} and \textsc{SevenNet}, training-from-scratch is at least competitive with fine-tuning and, in many cases, slightly more accurate.
With $2{,}000$~training configurations, force errors are often below $0.02$~eV/{\AA} and energy errors below $0.001$~eV/atom, whereas the corresponding fine-tuned models often show force errors around $0.03$~eV/{\AA} and similar energy errors.
Because these values approach the intrinsic uncertainty of the DFT reference data itself \cite{wang2021dft_error}, differences between fine-tuning and training-from-scratch should not be overinterpreted.
Nevertheless, judged by the numerical errors alone, training-from-scratch is frequently the more accurate strategy for \textsc{MACE} and \textsc{SevenNet} in the $200$- and $2{,}000$-configuration regimes.
In contrast, \textsc{GRACE} appears to require more data to reach comparable accuracy when trained-from-scratch.

The sulfur-vacancy jump again highlights that low trajectory-level errors are necessary but not sufficient.
Figure~\ref{fig:train}d compares the potential-energy curves predicted by fine-tuned \textsc{MACE-MP-0} models and similarly sized \textsc{MACE} models trained-from-scratch.
As in the fine-tuning case, only models trained on $2{,}000$~configurations recover the DFT energy profile.
However, the curves predicted by the models trained-from-scratch on $200$~and $2{,}000$~configurations are less smooth than the corresponding fine-tuned curves.
Because smoothness of the potential-energy surface can serve as a qualitative indicator of MLIP reliability, this observation complicates a simple ranking based only on force and energy MAEs.
For example, the trained-from-scratch \textsc{MACE} model has a force error of $0.013$~eV/{\AA}, compared with $0.030$~eV/{\AA} for the fine-tuned model and an energy error of $0.0007$~eV/atom, compared with $0.0011$~eV/atom.
Despite these lower numerical errors, the less smooth energy profile suggests that the fine-tuned model may still provide a more physically consistent reaction path in this specific case.

The comparison between fine-tuning and training-from-scratch also depends on the MLIP framework.
For \textsc{SevenNet}, only the model trained-from-scratch on $2{,}000$~configurations reproduces the correct shape of the energy curve (Supplementary Note~9).
For \textsc{GRACE}, none of the models trained-from-scratch recover the DFT energy profile in this benchmark (Supplementary Note~9).
The remaining material-specific properties are analyzed in Supplementary Notes~10 and~11 for trained-from-scratch \textsc{MACE} and \textsc{SevenNet} models, respectively.
Both frameworks reproduce the DFT reference properties with high accuracy across the investigated systems, although the hydroxide-ion mobility remains overestimated.

These results support a practical recommendation.
For material-specific MLIPs, a robust starting point is to train \textsc{MACE} or \textsc{SevenNet} from scratch on approximately $2{,}000$~configurations obtained by sub-sampling an AIMD trajectory using stride $10$ (see Supplementary Notes~12 and 13).
This corresponds to a trajectory of about $20{,}000$~frames.
For the smaller systems considered here, such trajectories can often be generated within one to two days; however, the cost increases substantially for systems containing more than $100$~atoms.

\subsection*{Model averaging in scarce-data regimes}

\begin{figure*}
    \centering
    \includegraphics{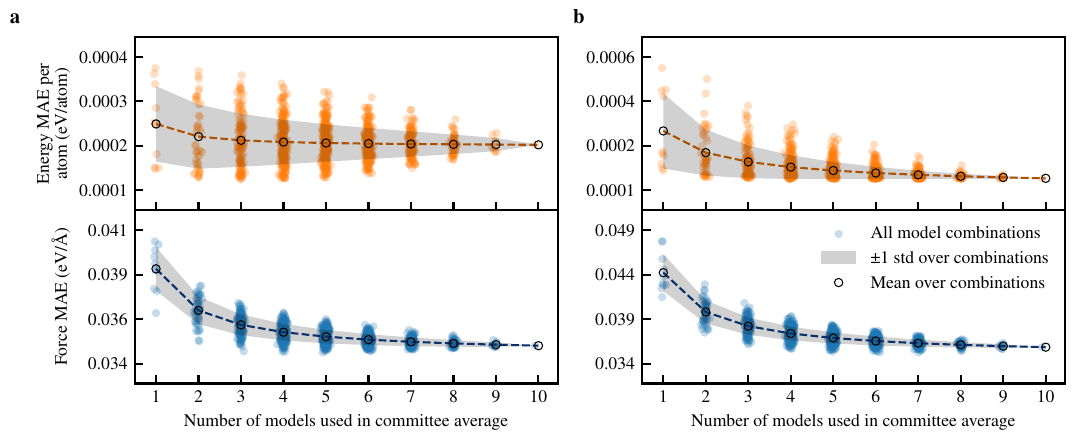}
    \caption{
    \textbf{Improving MLIP predictions in scarce-data regimes by model averaging.}
    Energy errors (top) and force errors (bottom) obtained by averaging MLIP outputs over all possible combinations of independently trained models.
    Results are shown for material-specific models of \textbf{a}~CDP and \textbf{b}~L-Pyro.
    The trained-from-scratch \textsc{MACE} models were trained on sub-sampled AIMD trajectories containing $200$~configurations using stride $10$.
    }
    \label{fig:model_avg}
\end{figure*}

When reference data are scarce, another way to improve predictions without generating additional DFT data is to train several models from scratch on the same dataset using different random seeds and to average their outputs.
This strategy is related to committee approaches commonly used to quantify MLIP uncertainty, but here we focus on its effect on predictive accuracy.
Considering all possible combinations of independently trained models, we find that increasing the number of averaged models reduces both the spread of the predictions and the mean error.
The improvement is most pronounced for force errors, which are often reduced by $10$-$20$~\% without adding any new training structures.
Most of the gain is already obtained by averaging two or three models, which typically improves the predictions by $8$-$10$~\%.
The average over all possible combinations of one to ten models trained with different random seeds is shown in Figure~\ref{fig:model_avg} for CDP and L-Pyro.

\section*{Discussion}

This work systematically examined how training-set size and dataset construction affect the generation of ab initio-accurate material-specific MLIPs.
We focused on two simple and broadly accessible strategies: naive fine-tuning of universal models and training models from scratch on system-specific reference data.
More advanced strategies, such as active learning, low-rank adaptation, or multi-head fine-tuning, were deliberately excluded.

For fine-tuning, the central result is that $2{,}000$~reference configurations sampled from AIMD trajectories provide a robust default for the investigated systems.
Across seven materials and five universal MLIPs, this dataset size consistently produced models with low force and energy errors and, more importantly, models that reproduced the relevant physical observables.
Datasets containing only $200$~configurations can be sufficient in favorable cases, but the outcome is less predictable and strongly system-dependent.

The manner in which AIMD trajectories are sub-sampled also matters, but within the tested regimes it matters less than the total number of training configurations.
Reducing the stride from $100$ to $10$ yields $2{,}000$~configurations from a trajectory that is ten times shorter.
Although the resulting structures are more strongly correlated, the models trained on these datasets perform comparably to those trained on more widely spaced configurations.
This suggests that moderately dense sampling is a cost-effective way to obtain a sufficient number of reference structures for material-specific MLIPs.
The stride-$1$ scaling study, however, also shows that overly dense sampling eventually introduces substantial redundancy:
for the MoS$_2$ sulfur-vacancy jump, dense datasets required at least $8{,}000$~configurations to reliably recover the DFT energy profile, whereas $2{,}000$~configurations sufficed when using stride $10$ or $100$.

Training-from-scratch provides a competitive alternative to fine-tuning.
For \textsc{MACE} and \textsc{SevenNet}, models trained-from-scratch on the same AIMD-derived datasets often performed on par with, or slightly better than, the corresponding fine-tuned models, most clearly in the $200$- and $2{,}000$-configuration regimes.
By contrast, \textsc{GRACE} required more data to reach comparable accuracy when trained-from-scratch.

The results also emphasize that trajectory-level force and energy errors alone are not sufficient to assess MLIP quality \cite{fu2022forces}.
The MoS$_2$ sulfur-vacancy jump demonstrates that models with apparently acceptable MAEs can still fail to reproduce a physically meaningful reaction profile.
For reactive events, migration barriers and other rare processes, direct validation against the target observable should therefore remain an essential part of MLIP assessment.

For practical applications, we recommend the following workflow.
First, generate an AIMD trajectory that samples the relevant material state and local environments.
Second, construct a training set of approximately $2{,}000$~configurations, for example by sub-sampling a $20{,}000$-frame trajectory using stride $10$.
Third, train material-specific \textsc{MACE} or \textsc{SevenNet} models from scratch, or alternatively fine-tune the corresponding universal models; in our benchmarks, training-from-scratch was often at least as accurate as naive fine-tuning when the same reference data were used.
Finally, when the highest reliability is needed and no additional DFT data are available, train several models with different random seeds and average their predictions.
This simple model-averaging strategy improves accuracy in scarce-data regimes without increasing the reference-data cost.

\section*{Methods}

\subsection*{AIMD simulations}

Reference DFT-based AIMD simulations were performed on CPUs using \textsc{CP2K} (version 2025.1) \cite{cp2k_1, cp2k_2, cp2k_3, cp2k_4, cp2k_5, cp2k_quickstep,cp2k_orb_trans} with the BLYP (KOH, L-Pyro, PhOH) \cite{blyp1,blyp2} and PBE (CDP, CPP, MoS$_2$, Li$_{13}$Si$_4$) \cite{pbe} exchange-correlation functionals, GTH pseudopotentials \cite{cp2k_gth-pseudopot1, cp2k_gth-pseudopot2, cp2k_gth-pseudopot3}, the DZVP-MOLOPT basis set \cite{cp2k_basis-set} and Nos\'{e}--Hoover chain thermostats \cite{nose1,nose2,nose3}.
A timestep of $0.5$~fs was used.
Training data consisted of $10$, $200$, or $2{,}000$~configurations per system, equidistantly sampled with strides of $10$ and $100$ from AIMD trajectories at the relevant temperatures: $300$~K (PhOH, L-Pyro), $333$~K (KOH), $500$~K (Li$_{13}$Si$_4$), [$510$, $540$, $585$, $620$, $660$~K] (CDP, CPP) and $1000$~K (MoS$_2$).

\subsection*{Training and fine-tuning MLIPs}

All fine-tuning and training calculations were performed using the official releases of the respective MLIP framework implementations: \textsc{MACE-torch} (version 0.3.14) \cite{mace_1,mace_2}, \textsc{GRACE tensorpotential} (version 0.5.1) \cite{grace_1,grace_2}, \textsc{SevenNet} (version 0.11.2) \cite{sevennet_1,sevennet_2}, \textsc{MatterSim} (version 1.1.2) \cite{mattersim} and \textsc{ORB} (version 0.3.2) \cite{orbv2,orbv3}.
For fine-tuning, we employed small universal models of the respective MLIP frameworks: \textsc{MACE-MP-0} (small) \cite{mace_mp}, \textsc{GRACE-1L-OAM} \cite{grace_2}, \textsc{SevenNet-0} \cite{sevennet_1}, \textsc{MatterSim-v1.0.0-5M} (MatterSim-Large) \cite{mattersim} and \textsc{ORB-v2} \cite{orbv2}.
For training-from-scratch, we used the same model sizes as these small universal models to enable a direct comparison between fine-tuning and training-from-scratch.
Fine-tuning and training were performed on \textsc{NVIDIA} A100 GPUs.

All errors reported in this work were computed on reference data excluded from the training and validation sets.
This test dataset was obtained by first-principles recalculation of the MLIP trajectories generated with fine-tuned universal \textsc{MACE-MP-0} models trained on $2{,}000$~data points from AIMD trajectories sub-sampled using stride $100$.

\subsection*{MLIP MD and NEB calculations}

Molecular dynamics simulations were performed with \textsc{ASE} (version 3.25.0) \cite{ase} using Nos\'{e}-Hoover chain thermostats \cite{nose1, nose2, nose3}, set up with the \textsc{aMACEing\_toolkit} (version 0.7.0) \cite{haenseroth2026atk}, for $3$~ns with a timestep of $0.5$~fs at $300$~K (PhOH, L-Pyro), $333$~K (KOH), $500$~K (Li$_{13}$Si$_4$), [$510$, $540$, $585$, $620$, $660$~K] (CDP, CPP) and $1000$~K (MoS$_2$).
Nudged elastic band calculations were performed with \textsc{ASE} (version 3.25.0) \cite{ase}.
All MLIP-MD and NEB calculations were performed on \textsc{NVIDIA} A100 GPUs.

\section*{Data Availability}

The dataset containing training sets and the models are available at \url{doi.org/10.5281/zenodo.21198813}.

\section*{Code Availability}

The used third-party codes \textsc{CP2K}, \textsc{MACE}, \textsc{MatterSim}, \textsc{GRACE}, \textsc{SevenNet}, \textsc{UPET} and \textsc{ORB} are available at \url{cp2k.org}, \url{github.com/acesuit/mace}, \url{github.com/microsoft/mattersim}, \url{github.com/ICAMS/grace-tensorpotential}, \url{github.com/MDIL-SNU/SevenNet}, \url{github.com/lab-cosmo/upet} and \url{github.com/orbital-materials/orb-models} respectively. 
The input-script creator used in this study is available at \url{github.com/jhaens/amaceing_toolkit}.

\section*{Acknowledgments}

We thank the staff of the Compute Center of the Technische Universit{\"a}t Ilmenau, especially Mr.~Henning~Schwanbeck, for providing an excellent research environment.
J.~H{\"a}nseroth thanks J.~L.~Wolf for help that contributed to the development of the committee approach used to reduce the force and energy errors shown in Figure~\ref{fig:model_avg} and P.~Merle for his valuable feedback on the first draft of the manuscript.
This work is supported by the doctoral scholarship of the German Academic Scholarship foundation, the Carl-Zeiss-Stiftung (SustEnMat, funding code: P2023-02-008), the Th{\"u}ringer Aufbaubank (TAB) (KapMemLyse, grant no.~2024 FGR 0081 / 0082), and the European Social Fund Plus (ESF+).

\section*{Competing interests}

The authors declare no competing interests.

\section*{Author contributions}
J.H. and C.D.~conceived the idea, J.H.~wrote the high-throughput workflow, performed all calculations; J.H. and C.D.~analyzed the data; J.H.~visualized all results and J.H.~wrote the first draft of the manuscript. 
C.D.~supervised the work; all authors revised and approved the manuscript.

\bibliography{bibliography.bib}

\end{document}


\author{Jonas H{\"a}nseroth}
\email{jonas.haenseroth@tu-ilmenau.de}
\affiliation{Theoretical Solid State Physics, Institute of Physics, Technische Universit{\"a}t Ilmenau, 98693 Ilmenau, Germany}

\author{Christian Dre{\ss}ler}
\affiliation{Theoretical Solid State Physics, Institute of Physics, Technische Universit{\"a}t Ilmenau, 98693 Ilmenau, Germany}

\title{Supplementary Information for "Data-Efficient Construction of Material-Specific Machine-Learning Interatomic Potentials from Ab Initio Molecular Dynamics Trajectories"}
\date{\today}

\maketitle

For more information on abbreviations, please refer to the main text, where all abbreviations are defined in detail.
Abbreviations not introduced in the main text are defined here.

\tableofcontents
\newpage
\clearpage

\newpage
\section*{Supplementary Note 1: Universal model performance - Force errors}

\begin{figure}[ht]
    \centering
    \includegraphics{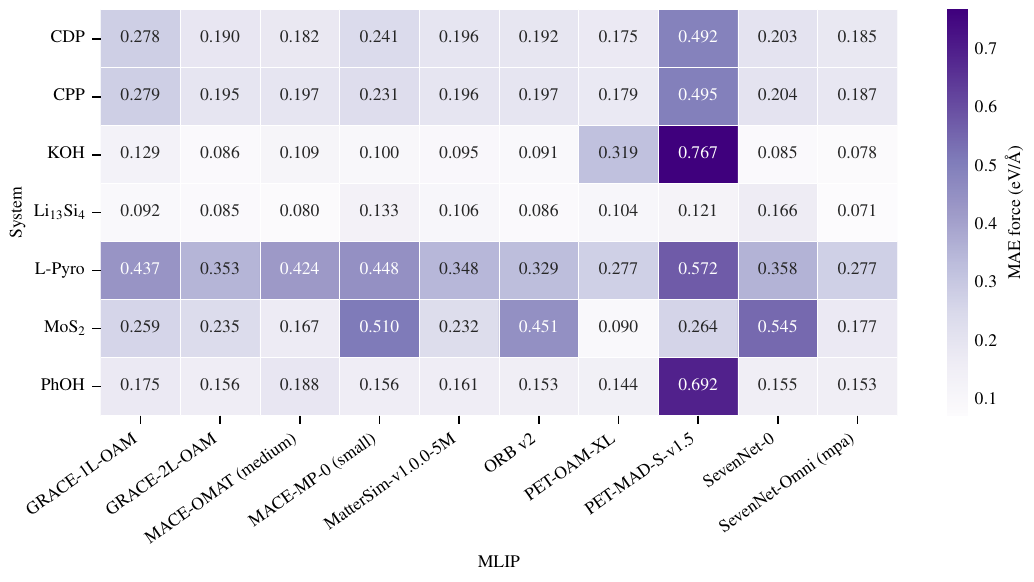}
    \caption{\textbf{Universal model force errors}.
    Mean-absolute-error on the predicted forces (in eV/\AA) of the different universal MLIPs on the different systems: CsH$_2$PO$_4$ (CDP) and Cs$_7$(H$_4$PO$_4$)(H$_2$PO$_4$)$_8$ (CPP), aqueous KOH solution, Li$_{13}$Si$_4$, MoS$_2$, phenol in water and L-pyroglutamate-ammonium (L-Pyro).
    } 
    \label{fig:found_model_frc_error}
\end{figure}

\newpage
\section*{Supplementary Note 2: Universal model performance - Energy errors}

\begin{figure}[ht]
    \centering
    \includegraphics{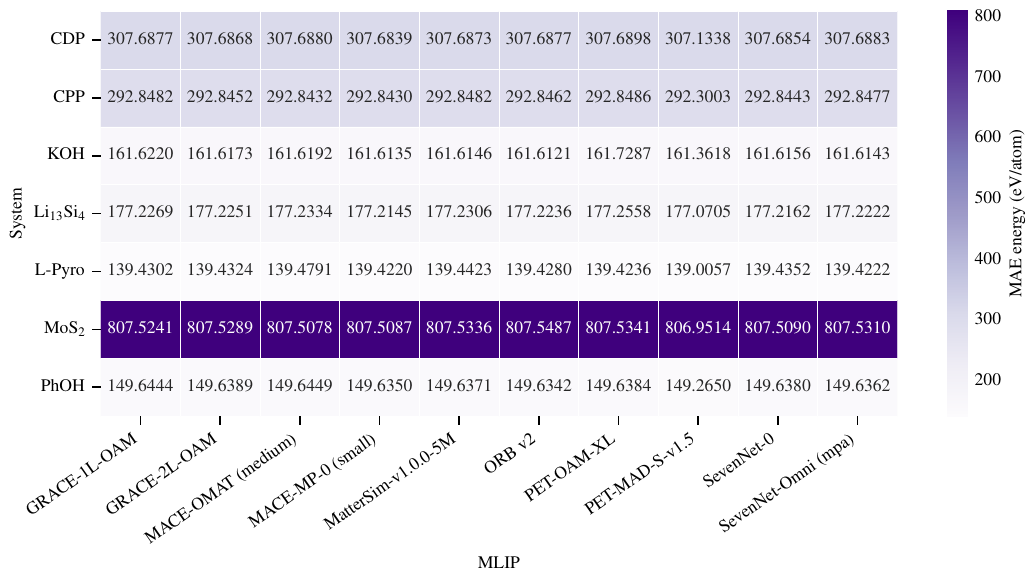}
    \caption{\textbf{Universal model energy errors}.
    Mean-absolute-error on the predicted energies of the different universal MLIPs on the different systems: CsH$_2$PO$_4$ (CDP) and Cs$_7$(H$_4$PO$_4$)(H$_2$PO$_4$)$_8$ (CPP), aqueous KOH solution, Li$_{13}$Si$_4$, MoS$_2$, phenol in water and L-pyroglutamate-ammonium (L-Pyro).
    } 
    \label{fig:found_model_ener_error}
\end{figure}

\newpage
\section*{Supplementary Note 3: Fine-tuned model performance - Energy errors - MLIPs vs. systems vs. dataset size (stride 100)}

\begin{figure}[ht]
    \centering
    \includegraphics{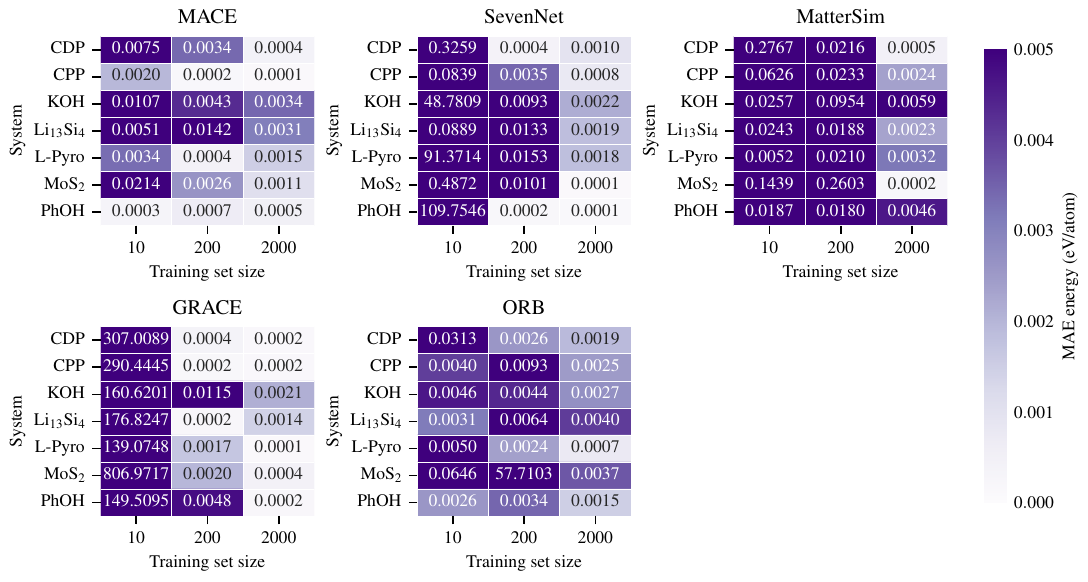}
    \caption{\textbf{Fine-tuned universal model energy errors}.
    Mean-absolute-error on the predicted energies of the different fine-tuned universal MLIPs (AIMD dataset, 2000 frames, stride 100) on the different systems: CsH$_2$PO$_4$ (CDP) and Cs$_7$(H$_4$PO$_4$)(H$_2$PO$_4$)$_8$ (CPP), aqueous KOH solution, Li$_{13}$Si$_4$, MoS$_2$, phenol in water and L-pyroglutamate-ammonium (L-Pyro).
    } 
    \label{fig:stride_100_ft_model_ener_error}
\end{figure}

\newpage
\section*{Supplementary Note 4: Fine-tuned universal model (AIMD dataset) - prediction of potential energy profile for sulfur-vacancy jump in MoS$_2$}

\begin{figure}[ht]
    \centering
    \includegraphics{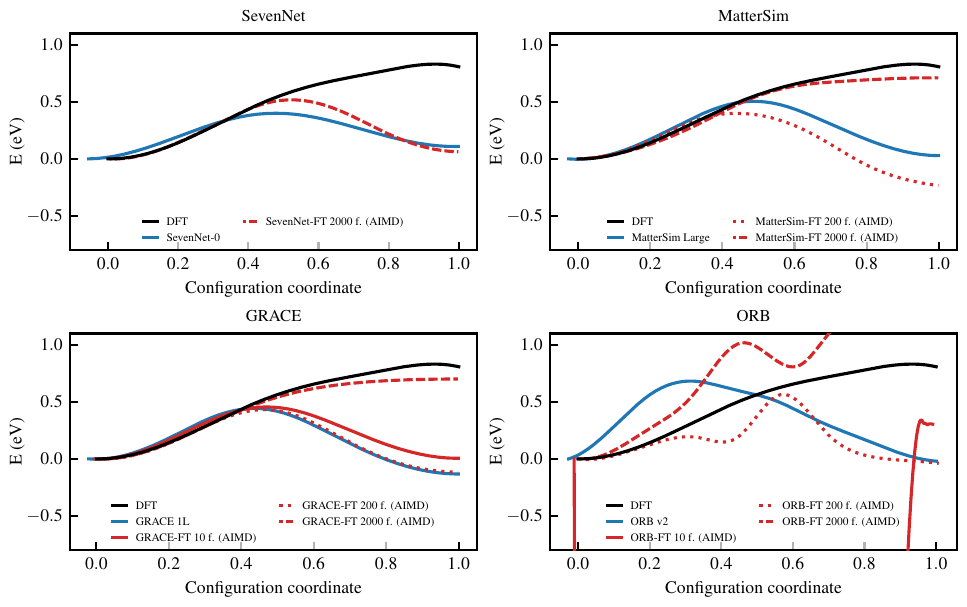}
    \caption{\textbf{Energy profiles computed with fine-tuned universal models with variable AIMD dataset size}.
    Potential energy curves for a sulfur jump into a neighboring line of sulfur vacancies in MoS$_2$ via NEB calculations using fine-tuned universal models with training datasets of size $10$~(solid red lines), $200$~(dotted red lines) and $2000$~(dashed red lines) and the respective universal model (solid blue line) as well as the DFT reference profile (solid black line).
    The universal \textsc{SevenNet-0} models fine-tuned on the datasets consisting out of $10$~and $200$~data points sub-sampled from AIMD trajectories as well as the universal \textsc{MatterSim-v1.0.0-5M} model fine-tuned on the dataset consisting out of $10$~data points sub-sampled from AIMD trajectories are not able to compute converged structures for the NEB replicas.
    The minimum of the energy profile obtained with the fine-tuned \textsc{ORB-v2} model using the dataset consisting out of $10$~data points is $-121.25$~eV.
    The maximum of the energy profile obtained with the fine-tuned \textsc{ORB-v2} model using the dataset consisting out of $2000$~data points is $1.38$~eV.
    }
    \label{fig:ft_aimd_mos2}
\end{figure}

\newpage
\section*{Supplementary Note 5: MACE-MP-0 fine-tuned universal model (AIMD dataset) - prediction of physical properties}

\begin{figure}[ht]
    \centering
    \includegraphics{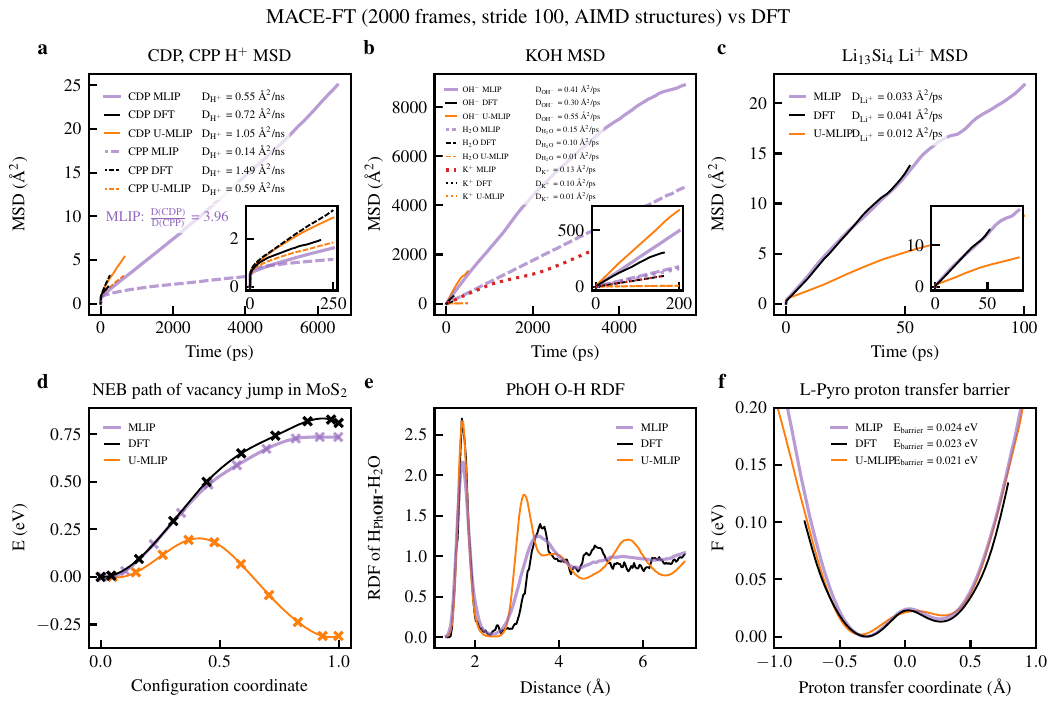}
    \caption{\textbf{MACE-MP-0 finetuned universal model on AIMD dataset}.
    Comparison of different physical properties obtained with first principles methods and universal model MACE-MP-0 and the fine-tuned universal models (AIMD dataset, 2000 frames, stride 100): \textbf{a} CsH$_2$PO$_4$ (CDP) and Cs$_7$(H$_4$PO$_4$)(H$_2$PO$_4$)$_8$ (CPP) - proton diffusion coefficients ratios of D(CDP)/D(CPP), \textbf{b} KOH - water, water molecule and hydroxide ion mean-squared displacements and diffusion coefficients, \textbf{c} Li$_{13}$Si$_4$ - lithium ion mean-squared displacements and diffusion coefficients, \textbf{d} MoS$_2$ - potential energy curves for a sulfur jump into a neighboring line of sulfur vacancies, \textbf{e} phenol in water - (H$_2$O)$\cdots$O\textsubscript{Hydroxyl-Group} radial distribution function and \textbf{f} L-pyroglutamate-ammonium (L-Pyro) - free energy profiles along the proton transfer coordinate.
    } 
    \label{fig:mace_ft_aimd_2000_100}
\end{figure}

\newpage
\section*{Supplementary Note 6: Fine-tuned model performance - Force errors - MLIPs vs. systems vs. dataset size (stride 10)}

\begin{figure}[ht]
    \centering
    \includegraphics{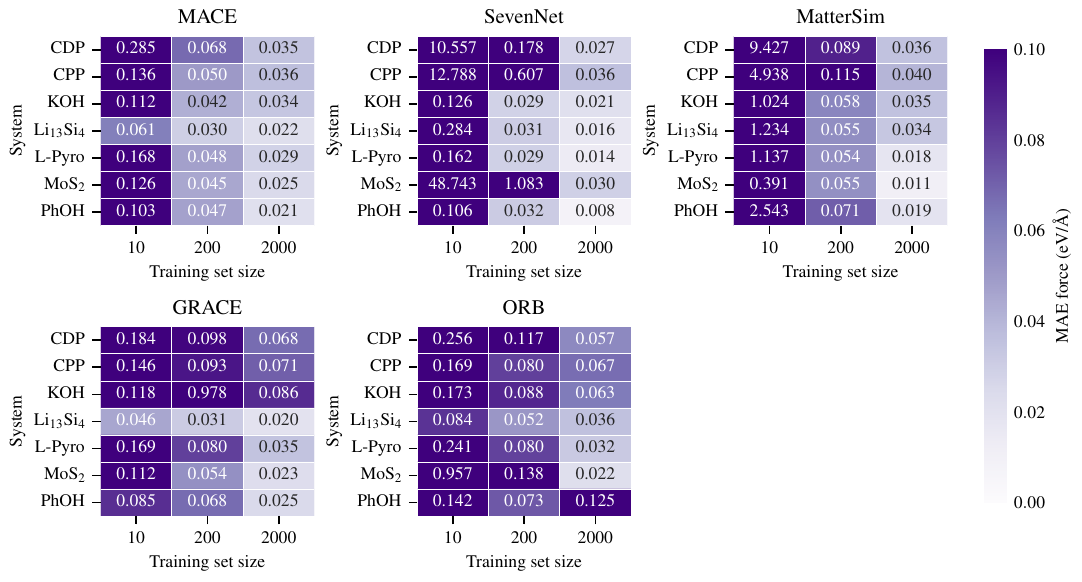}
    \caption{\textbf{Fine-tuned universal model energy errors}.
    Mean-absolute-error on the predicted forces of the different fine-tuned universal MLIPs (AIMD dataset, 2000 frames, stride 10) on the different systems: CsH$_2$PO$_4$ (CDP) and Cs$_7$(H$_4$PO$_4$)(H$_2$PO$_4$)$_8$ (CPP), aqueous KOH solution, Li$_{13}$Si$_4$, MoS$_2$, phenol in water and L-pyroglutamate-ammonium (L-Pyro).
    } 
    \label{fig:stride_10_ft_model_ener_error}
\end{figure}

\newpage
\section*{Supplementary Note 7: Fine-tuned model performance - Energy errors - MLIPs vs. systems vs. dataset size (stride 10)}

\begin{figure}[ht]
    \centering
    \includegraphics{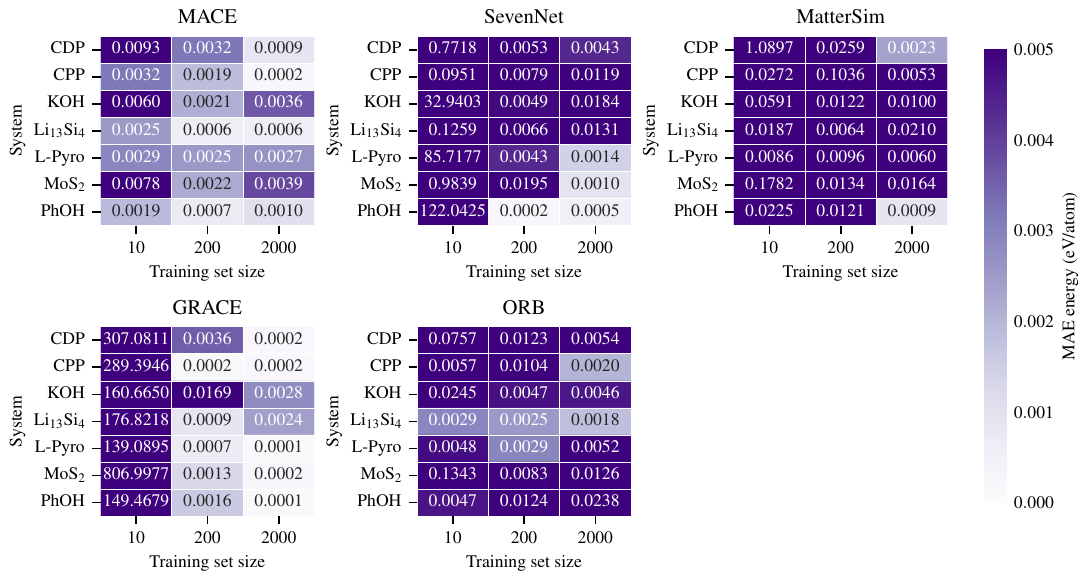}
    \caption{\textbf{Fine-tuned universal model energy errors}.
    Mean-absolute-error on the predicted energies of the different fine-tuned universal MLIPs (AIMD dataset, 2000 frames, stride 10) on the different systems: CsH$_2$PO$_4$ (CDP) and Cs$_7$(H$_4$PO$_4$)(H$_2$PO$_4$)$_8$ (CPP), aqueous KOH solution, Li$_{13}$Si$_4$, MoS$_2$, phenol in water and L-pyroglutamate-ammonium (L-Pyro).
    } 
    \label{fig:stride_10_ft_model_ener_error}
\end{figure}

\newpage
\section*{Supplementary Note 8: Trained model performance - Energy errors - MLIPs vs. systems vs. dataset size (stride 100)}

\begin{figure}[ht]
    \centering
    \includegraphics{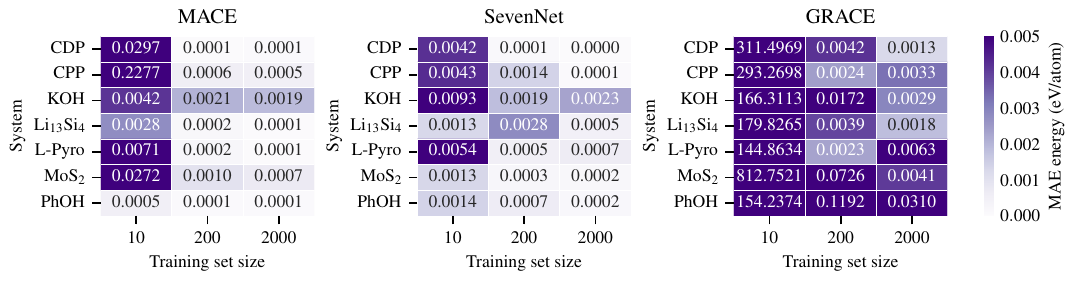}
    \caption{\textbf{Trained-from-scratch model energy errors}.
    Mean-absolute-error on the predicted energies of the different trained-from-scratch MLIPs (AIMD dataset, 2000 frames, stride 100) on the different systems: CsH$_2$PO$_4$ (CDP) and Cs$_7$(H$_4$PO$_4$)(H$_2$PO$_4$)$_8$ (CPP), aqueous KOH solution, Li$_{13}$Si$_4$, MoS$_2$, phenol in water and L-pyroglutamate-ammonium (L-Pyro).
    } 
    \label{fig:stride_100_train_model_ener_error}
\end{figure}

\newpage
\section*{Supplementary Note 9: Trained-from-scratch model (AIMD dataset) - prediction of potential energy profile for sulfur-vacancy jump in MoS$_2$}

\begin{figure}[ht]
    \centering
    \includegraphics{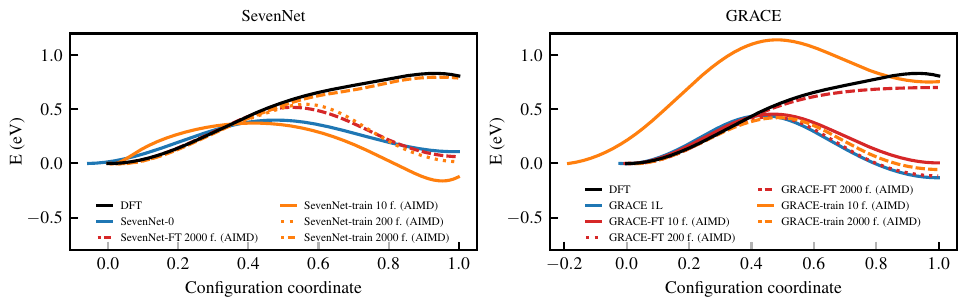}
    \caption{\textbf{Energy profiles computed with trained models with variable AIMD dataset size}.
    Potential energy curves for a sulfur jump into a neighboring line of sulfur vacancies in MoS$_2$ via NEB calculations using fine-tuned (red lines) and trained-from-scratch (orange) models: \textsc{SevenNet} (left) and \textsc{GRACE} (right); with data set sizes of $10$~(solid lines), $200$~(dotted lines) and $2000$~(dashed lines) and the respective universal model (solid blue line) as well as the DFT reference profile (solid black line).
    The universal \textsc{SevenNet-0} models fine-tuned on the datasets consisting out of $10$~and $200$~data points sub-sampled from AIMD trajectories as well as the \textsc{GRACE} model trained-from-scratch on the dataset consisting out of $200$~data points sub-sampled from AIMD trajectories are not able to compute converged structures for the NEB replicas.
    } 
    \label{fig:train_vs_ft_aimd_mos2}
\end{figure}

\newpage
\section*{Supplementary Note 10: MACE trained-from-scratch model (AIMD dataset) - prediction of physical properties}

\begin{figure}[ht]
    \centering
    \includegraphics{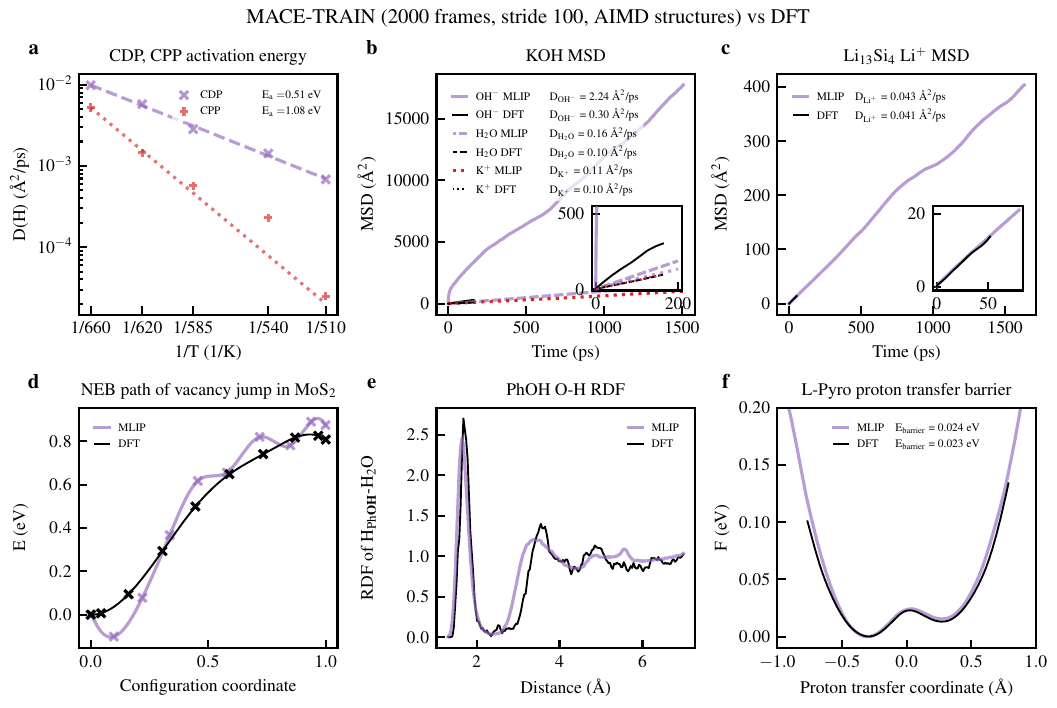}
    \caption{\textbf{MACE trained on AIMD dataset}.
    Comparison of different physical properties obtained with first principles methods and the trained-from-scratch models (AIMD dataset, 2000 frames, stride 3000): \textbf{a} CsH$_2$PO$_4$ (CDP) and Cs$_7$(H$_4$PO$_4$)(H$_2$PO$_4$)$_8$ (CPP) - proton diffusion coefficients ratios of D(CDP)/D(CPP), \textbf{b} KOH - water, water molecule and hydroxide ion mean-squared displacements and diffusion coefficients, \textbf{c} Li$_{13}$Si$_4$ - lithium ion mean-squared displacements and diffusion coefficients, \textbf{d} MoS$_2$ - potential energy curves for a sulfur jump into a neighboring line of sulfur vacancies, \textbf{e} phenol in water - (H$_2$O)$\cdots$O\textsubscript{Hydroxyl-Group} radial distribution function and \textbf{f} L-pyroglutamate-ammonium (L-Pyro) - free energy profiles along the proton transfer coordinate.
    } 
    \label{fig:mace-train}
\end{figure}

\newpage
\section*{Supplementary Note 11: SevenNet trained-from-scratch model (AIMD dataset) - prediction of physical properties}

\begin{figure}[ht]
    \centering
    \includegraphics{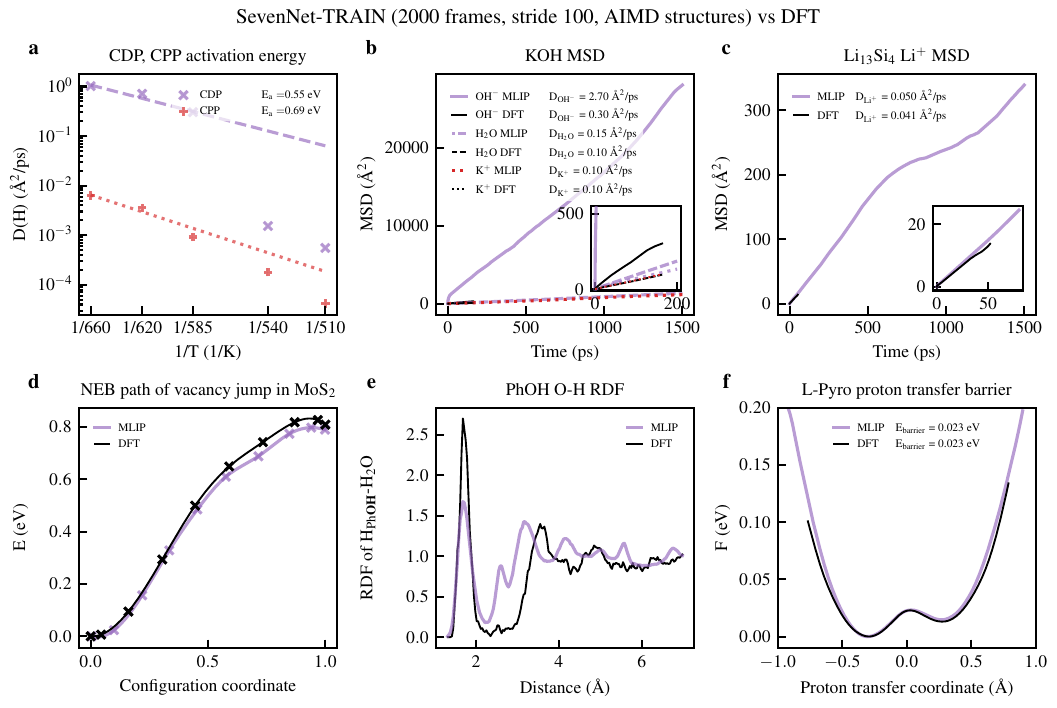}
    \caption{\textbf{SevenNet trained on AIMD dataset}.
    Comparison of different physical properties obtained with first principles methods and the trained-from-scratch models (AIMD dataset, 2000 frames, stride 3000): \textbf{a} CsH$_2$PO$_4$ (CDP) and Cs$_7$(H$_4$PO$_4$)(H$_2$PO$_4$)$_8$ (CPP) - proton diffusion coefficients ratios of D(CDP)/D(CPP), \textbf{b} KOH - water, water molecule and hydroxide ion mean-squared displacements and diffusion coefficients, \textbf{c} Li$_{13}$Si$_4$ - lithium ion mean-squared displacements and diffusion coefficients, \textbf{d} MoS$_2$ - potential energy curves for a sulfur jump into a neighboring line of sulfur vacancies, \textbf{e} phenol in water - (H$_2$O)$\cdots$O\textsubscript{Hydroxyl-Group} radial distribution function and \textbf{f} L-pyroglutamate-ammonium (L-Pyro) - free energy profiles along the proton transfer coordinate.
    } 
    \label{fig:sevennet-train}
\end{figure}

\newpage
\section*{Supplementary Note 12: Trained model performance - Force errors - MLIPs vs. systems vs. dataset size (stride 10)}

\begin{figure}[ht]
    \centering
    \includegraphics{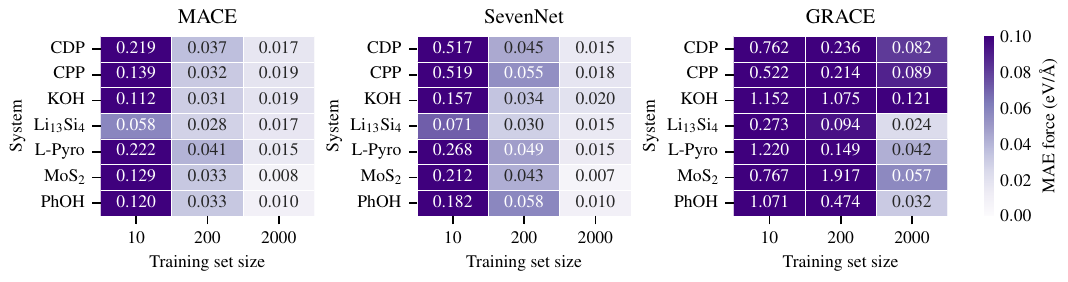}
    \caption{\textbf{Trained-from-scratch model force errors}.
    Mean-absolute-error on the predicted forces of the different trained-from-scratch MLIPs (AIMD dataset, 2000 frames, stride 10) on the different systems: CsH$_2$PO$_4$ (CDP) and Cs$_7$(H$_4$PO$_4$)(H$_2$PO$_4$)$_8$ (CPP), aqueous KOH solution, Li$_{13}$Si$_4$, MoS$_2$, phenol in water and L-pyroglutamate-ammonium (L-Pyro).
    } 
    \label{fig:stride_10_train_model_frc_error}
\end{figure}

\newpage
\section*{Supplementary Note 13: Trained model performance - Energy errors - MLIPs vs. systems vs. dataset size (stride 10)}

\begin{figure}[ht]
    \centering
    \includegraphics{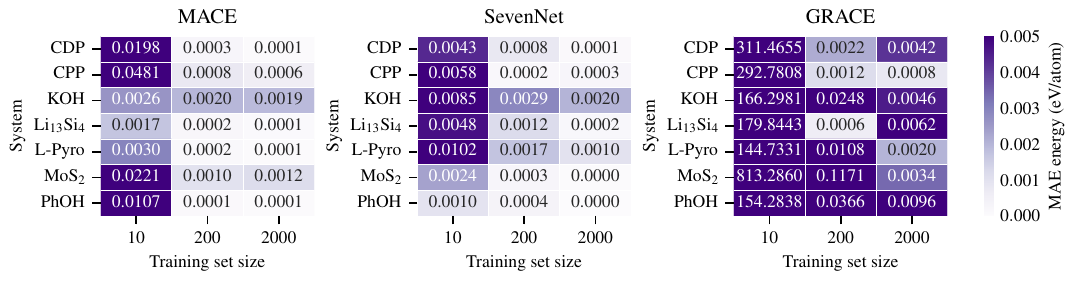}
    \caption{\textbf{Trained-from-scratch model energy errors}.
    Mean-absolute-error on the predicted energies of the different trained-from-scratch MLIPs (AIMD dataset, 2000 frames, stride 10) on the different systems: CsH$_2$PO$_4$ (CDP) and Cs$_7$(H$_4$PO$_4$)(H$_2$PO$_4$)$_8$ (CPP), aqueous KOH solution, Li$_{13}$Si$_4$, MoS$_2$, phenol in water and L-pyroglutamate-ammonium (L-Pyro).
    } 
    \label{fig:stride_10_train_model_ener_error}
\end{figure}